\documentclass[twocolumn,floats,floatfix,showpacs,amssymb,prd,superscriptaddress,nofootinbib,eqsecnum]{revtex4-1}
\usepackage{graphicx, epsfig, amssymb} 
\usepackage{amsmath, amsfonts,amsthm,latexsym}
\usepackage{bm} 

\usepackage[linktocpage]{hyperref}
\usepackage[caption=false]{subfig}
\usepackage[usenames]{color}

\usepackage{tabularx}
\usepackage{array}

\begin{document}

\title{\large The Casimir spectrum revisited}

\author{Carlos A.R.\ Herdeiro}\email{herdeiro@ua.pt}
\affiliation{Departamento de F\'\i sica da Universidade de Aveiro, 
Campus de Santiago, 3810-183 Aveiro, Portugal.}
\affiliation{Centro de F\'\i sica do Porto --- CFP,
Departamento de F\'\i sica e Astronomia,
Faculdade de Ci\^encias da Universidade do Porto --- FCUP,
Rua do Campo Alegre, 4169-007 Porto, Portugal.}

\author{Marco O.P.\ Sampaio}\email{msampaio@fc.up.pt}
\affiliation{Centro de F\'\i sica do Porto --- CFP,
Departamento de F\'\i sica e Astronomia,
Faculdade de Ci\^encias da Universidade do Porto --- FCUP,
Rua do Campo Alegre, 4169-007 Porto, Portugal.}

\author{Jaime E. Santos}\email{jesantos@pks.mpg.de}
\affiliation{Centro de F\'\i sica do Porto --- CFP,
Departamento de F\'\i sica e Astronomia,
Faculdade de Ci\^encias da Universidade do Porto --- FCUP,
Rua do Campo Alegre, 4169-007 Porto, Portugal.}
\affiliation{Max Planck Institute for the Physics of Complex Systems, \\
Noethnitzer Str. 38, D-01187 Dresden, Germany.}

\begin{abstract}

We examine the mathematical and physical significance of the spectral density $\sigma(\omega)$ introduced by Ford in~\cite{Ford:1988gt}, defining the contribution of each frequency to the \textit{renormalised} energy density of a quantum field. Firstly, by considering a simple example, we argue that $\sigma(\omega)$ is well defined, in the sense of being regulator independent, despite an apparently regulator dependent definition. We then suggest that $\sigma(\omega)$ is a spectral distribution, rather than a function, which only produces physically meaningful results when integrated over a sufficiently large range of frequencies and with a high energy smooth enough regulator. Moreover, $\sigma(\omega)$ is seen to be simply the difference between the bare spectral density and the spectral density of the reference background. This interpretation yields a simple `rule of thumb' to writing down a (formal) expression for $\sigma(\omega)$ as shown in an explicit example. Finally, by considering an example in which the sign of the Casimir force varies, we show that the spectrum carries no manifest information about this sign; it can only be inferred by integrating $\sigma(\omega)$.

\end{abstract}

\maketitle

\section{Introduction}
Can one assign a frequency spectrum to the finite Casimir energy? This question was first posed by Ford in~\cite{Ford:1988gt} where such frequency spectrum, henceforth dubbed \textit{Casimir spectrum}, was considered for two examples: a massless scalar field in $\mathbb{R}\times S^1$ and $\mathbb{R}^3\times S^1$. The spectrum is encoded in a spectral density $\sigma(\omega)$ which must obey two obvious criteria: i) integrated over all frequencies it must yield the renormalised vacuum energy
\begin{equation}
\rho_{ren}\equiv \langle 0| T_{00}|0\rangle_{ren}=\frac{1}{2\pi}\int_0^{+\infty} \sigma(\omega) d\omega \ , \label{maindef} \end{equation}
where the quantum field has energy momentum tensor $T_{\mu\nu}$ in a space-time admitting a globally time-like Killing vector field $\partial/\partial_{x^0}$; ii) it should not depend on the renormalisation/regularisation procedure. Whereas the first criterion is easy to check, the second one seems hard to prove. Indeed, the very definition of the spectral density introduced in~\cite{Ford:1988gt} was based on a specific regularisation/renormalisation scheme, namely point splitting. This definition  is
\begin{equation}
\sigma(\omega)=2\int_{-\infty}^{+\infty} \langle 0|T_{00}(\tau)|0\rangle_{ren} e^{i\omega \tau} d\tau \ ,
\label{sf}
 \end{equation}
where $\tau$ is a point-splitting regulator along the time component. We shall show in Section \ref{shortcut}, however, that this definition, albeit apparently tied to point-splitting regularisation, is quite general, and thus, that in this respect the Casimir spectrum is well defined. From the same analysis we shall learn that renormalised spectral density $\sigma(\omega)$ is simply the difference between the bare spectral density and the spectral density of the reference background. This provides us with an immediate tool to write down a formal expression for it. But if an appropriate regulator is not included, this expression is meaningless. Moreover, we shall see that even with an appropriate regulator, only the integration of this spectral density carries some physical information, and only if the integration extends through a sufficiently large frequency interval. Thus one should regard $\sigma(\omega)$ as a spectral \textit{distribution}, rather than a function. 

As another test on the physical information carried by the Casimir spectrum, we analyse the case of a scalar field, with mass parameter and coupling to the curvature, in an Einstein Static Universe. It has been shown in~\cite{Herdeiro:2005zj,Herdeiro:2007eb}, that the renormalised energy momentum tensor associated to the vacuum fluctuations may obey or violate the strong energy condition, depending on the various parameters, therefore potentially sourcing repulsive gravity. Our analysis shows, however, than no hint about the sign of the Casimir force may be obtained, in this case, from the Casimir spectrum. Thus, although we find that the Casimir spectrum is mathematically well defined, we find no evidence about it carrying more physical information than that carried by the Casimir energy itself.

This paper is organised as follows. In section 2 we start by reviewing the computation of the Casimir spectrum for a massless scalar field in $\mathbb{R}\times S^1$, discussing the generality of the definition \eqref{sf} as well as its interpretation and physical content. In section 3 we discuss the Casimir spectrum on the Einstein Static Universe, where we show that no noticeable difference is seen when the gravitational effect of the Casimir energy varies from attractive to repulsive. In Appendix~\ref{Check_Ford} we consider another example, a scalar field in $\mathbb{R}^3\times S^1$, that confirms the arguments and interpretation provided in Section 2.

\section{Reconsidering $ \mathbb{R}\times S^1$}
We start by reconsidering the case of a massless scalar field on a 2-dimensional space-time with one periodic space-like direction, of length $L$. The space-time is then $ \mathbb{R}\times S^1$. We shall review the computation of the spectrum using a weight function~\cite{Ford:1988gt,Lang_2005} and then compute it using point-splitting.

\subsection{Weight functions}
The regularised energy density (with a weight function) is
\begin{equation}
\rho_{x_0,m}=\dfrac{1}{2L}\sum_{n=-\infty}^{+\infty}\omega_n W(\omega_n)_{x_0,m} \ , 
\end{equation}
with 
$\omega_n=2\pi |n| /L$. We can apply the Abel-Plana formula as long as the condition
\[
\lim_{y\rightarrow\infty}e^{-2\pi y}\left|G(x\pm iy)\right|=0 \ ,
\] is satisfied (where $G(n)$ is the function inside the sum). The choice of weight function in~\cite{Ford:1988gt} is
\begin{equation}
W(2\pi x/L)_{x_0,m}=\left(\dfrac{2m}{x_0}\right)^{2m+1}\dfrac{x^{2m}}{(2m)!}e^{-2mx/x_0} \ , \label{wf}
\end{equation}
where $m\in \mathbb{N}$. This clearly obeys the conditions of applicability of the Abel-Plana formula.
Applying the formula we get
\begin{multline}
\rho_{x_0,m}=\dfrac{1}{2\pi}\int_0^{+\infty}\omega W(\omega)d\omega \\-\dfrac{2\pi}{L^2}\int_0^{+\infty}\dfrac{t\left(W(2\pi i t/L)+W(-2\pi i t/L\right))}{e^{2\pi t}-1}dt \; .
\end{multline}
The first term renormalises the cosmological constant since it is a background independent constant. The second term is identified as the spectral function. Note that the weight functions \eqref{wf} were chosen as to become delta functions $\delta(x-x_0)$ in the large $m$ limit, which when integrated over $x_0$ give 1. Also, $x_0=L\omega_0/(2\pi)$; the spectral function so defined is then normalised to give the correct energy density when integrated with this differential. Replacing by the explicit form of the weight function we get
\begin{multline}
\sigma_m(x_0)=\frac{2\pi(-1)^{m+1}}{L^2(2m)!}\left(\dfrac{2m}{x_0}\right)^{2m+1}\times \\ \times \int_0^{+\infty}t^{2m+1}\frac{\left(e^{\frac{2m i t}{x_0}}+e^{-\frac{2m i t}{x_0}}\right)}{e^{2\pi t}-1}dt \; .
\end{multline}
If we expand the denominator as a geometric series and perform the integrals, then the sums can be identified as Hurwitz zeta functions and the final expression coincides with that of~\cite{Lang_2005}:
\begin{multline}
\sigma_m(x_0)=\dfrac{(-1)^{m+1}(2m+1)}{L^2}\left(\dfrac{m}{\pi x_0}\right)^{2m+1}\times \\ \left[\zeta_H\left(2m+2,1-i\dfrac{m}{\pi x_0}\right)+\zeta_H\left(2m+2,1+i\dfrac{m}{\pi x_0}\right)\right] \; .
\end{multline}

\begin{figure*}
\begin{center}
\begin{picture}(0,80)
\put(0,80){\scriptsize $L^2\sigma_{1000}(x_0)$}
\put(250,80){\scriptsize $L^2\sigma_{10000}(x_0)$}
\end{picture}
\begin{tabular}{cc}
 \hspace{-4mm} \includegraphics[width=0.45\textwidth]{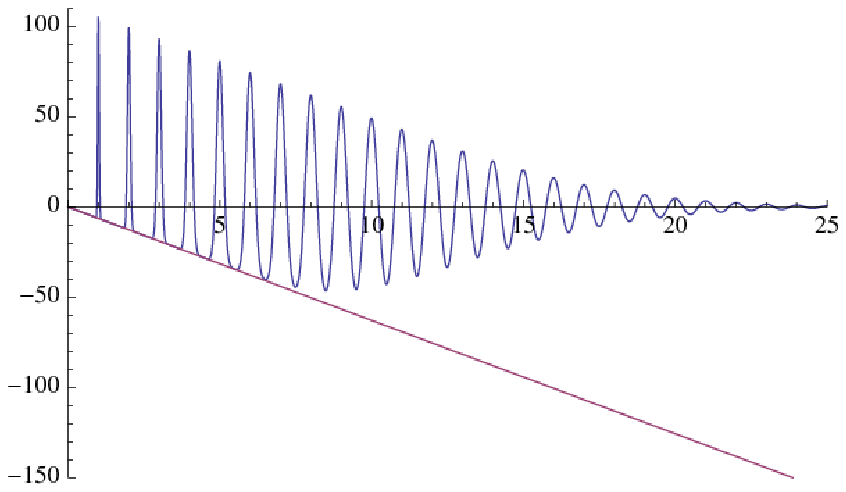} \ \ \ \ \  &
  \includegraphics[width=0.45\textwidth]{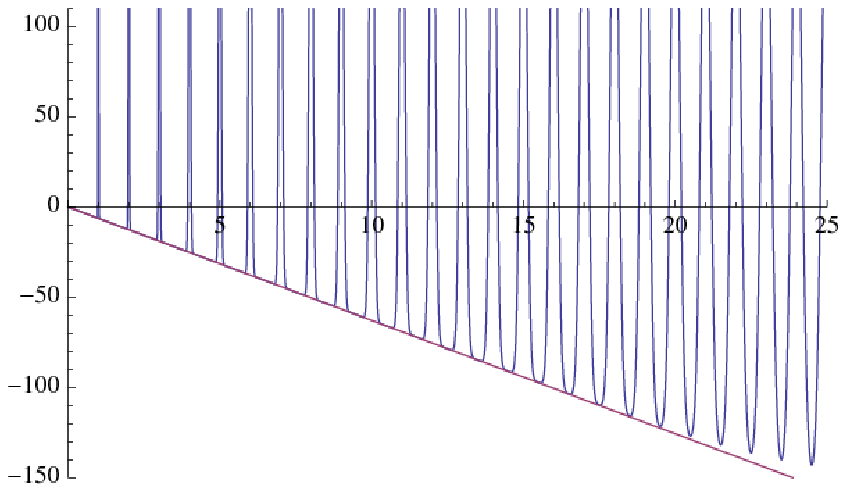}
\end{tabular}
\end{center}
   \caption{\label{Plots_weights} {\em Left:} Spectral function, $\sigma_m(x_0)$ from weight functions for $m=1000$, {\em Right:} Spectral function from weight functions for $m=10000$.} 
\end{figure*}

 The plot in Fig.~\ref{Plots_weights} reveals that in the limit of large $m$ the spectrum becomes just the difference between localised delta functions and a straight line with negative slope which matches the flat space spectrum.
Indeed,  re-writing the energy density as
\begin{eqnarray}
\rho &=& \dfrac{1}{2L}\sum_{n=-\infty}^{+\infty}\omega_n =\int_{0}^{+\infty} d\omega \omega\dfrac{1}{2L}\sum_{n=-\infty}^{+\infty}\delta(\omega-\omega_n) \nonumber\\ 
     &=&\dfrac{1}{L}\int_{0}^{+\infty} d\omega \omega\sum_{n=1}^{+\infty}\delta(\omega-\omega_n) 
\end{eqnarray}
and subtracting the flat space contribution (assuming a massless field), we get
\begin{eqnarray}
\rho_{ren} &=& \dfrac{1}{L}\int_{0}^{+\infty} d\omega \omega\sum_{n=1}^{+\infty}\delta(\omega-\omega_n)-\dfrac{1}{2}\int_{-\infty}^{+\infty}\dfrac{dk}{2\pi} \omega_k \nonumber \\
&=& \int_{0}^{+\infty} d\omega \omega\left(\sum_{n=1}^{+\infty}\dfrac{1}{L}\delta(\omega-\omega_n)-\dfrac{1}{2\pi}\right) \ .
\end{eqnarray}
Thus, in this simple case, it is clear that the spectral density is really
\begin{eqnarray}
\sigma(\omega) &=& \omega \left(2\pi\sum_{n=1}^{+\infty}\dfrac{1}{L}\delta(\omega-\omega_n)-1\right) \ .
\label{sigmas1} 
\end{eqnarray}
The terms in parenthesis show that the continuum frequencies have a negative uniform weight, whereas the discrete frequencies have a delta function uniform weight. The difference corresponds exactly to the limit of $m\rightarrow \infty$ in Fig.~\ref{Plots_weights}.

\subsection{Point Splitting}
The first form of the spectrum obtained in the previous section, depended on a particular choice of weight function. But the same result is obtained with point splitting regularisation (or any other well defined regularisation). To show this, let us first define the point split energy density. As shown in~\cite{Herdeiro:2007eb} the weight function is a cosine
\begin{equation}
\rho_{\tau}\equiv \langle 0|T_{00}(\tau)|0\rangle=\dfrac{1}{2L}\sum_{n=-\infty}^{+\infty}\omega_n \cos(\omega_n \tau) \ . \label{ps}
\end{equation}
Now, care must be take when applying the Abel-Plana formula. When the argument of the cosine is imaginary there is an exponentially growing part. Such term comes from the substitution $e^{-i 2\pi n \tau }\rightarrow e^{ -i 2\pi (x+iy)\tau } $, which marginally violates the convergence condition for the Abel-Plana formula. We can cure this by adding a small imaginary part to $\tau$ (which amounts to adding an exponentially decaying part to the cosine). If we multiply by a factor of the form $e^{-\epsilon\omega_n}$, and expand the cosine in~\eqref{ps} in terms of exponentials, then we have
\begin{eqnarray}
\rho_{\tau}&=&\dfrac{1}{2L}\sum_{n=0}^{+\infty}\omega_n \left(e^{i\omega_n(\tau+i\epsilon)}+e^{-i\omega_n(\tau-i\epsilon)}\right) \ . 
\end{eqnarray}
This gives a plus $\epsilon$ prescription for one of the exponentials in the cosine and a minus $\epsilon$ prescription for the other one. Applying now the Abel-Plana formula yields
\begin{widetext}
\begin{equation}
\rho_{\tau,ren}=-\dfrac{\pi}{L^2}\int_{0}^{\infty} dx\, \dfrac{x\left[e^{-2\frac{\pi}{L} x(\tau+i\epsilon)}+e^{2\frac{\pi}{L} x(\tau-i\epsilon)}+e^{2\frac{\pi}{L} x(\tau+i\epsilon)}+e^{-2\frac{\pi}{L} x(\tau-i\epsilon)}\right]}{e^{2\pi x}-1} \ ,
\end{equation}
\end{widetext}
where we have discarded the term 
\begin{equation}
\dfrac{1}{2\pi L}\int_0^{+\infty} d\omega \omega\cos(\omega \tau)e^{-\epsilon\omega} \ ,
\end{equation}
which amounts to performing the renormalisation of this quantity. For $\tau>1$ the Abel-Plana formula does not converge, however we know that the original sum converges for all $\tau$, so if we obtain an analytic expression when assuming $\tau<1$, it must hold for all $\tau$ because of the analyticity of the original sum. If we expand the denominator using the geometric series for $1/(1-e^{-2\pi x})$ and integrate term by term, we get
\begin{multline}
\rho_{\tau,ren}=-\dfrac{1}{4\pi L^2}\sum_{k=1}^{+\infty}\, \left[\dfrac{1 }{(k+\frac{\tau+i\epsilon}{L})^2}+\dfrac{1}{(k-\frac{\tau-i\epsilon}{L})^2}\right.+\\\left.+\dfrac{1}{(k-\frac{\tau+i\epsilon}{L})^2}+\dfrac{1}{(k+\frac{\tau-i\epsilon}{L})^2}\right]   .
\end{multline}
The definition of the spectral function~\eqref{sf} yields, 
\begin{multline}
\sigma(\omega)=-\dfrac{1}{2\pi L^2}\sum_{k=1}^{+\infty}\int_{-\infty}^{+\infty} e^{i\omega \tau}\left[\dfrac{1}{(k+\frac{\tau+i\epsilon}{L})^2}+\right. \\ \left.+\dfrac{1}{(k-\frac{\tau-i\epsilon}{L})^2}+\dfrac{1}{(k-\frac{\tau+i\epsilon}{L})^2}+\dfrac{1}{(k+\frac{\tau-i\epsilon}{L})^2}\right]d\tau \ .
\end{multline}
The poles of the integrand are above or below the real axis for $\mp i \epsilon$ respectively. We can close the contour on the upper part of the complex plane, because of the exponential which will give a vanishingly small contribution from the circle at infinity, so we reduce the problem to the calculation of residues. Then, after some manipulations we get
\begin{eqnarray}
\sigma(\omega)&=&2\omega\sum_{k=1}^{+\infty} e^{-\epsilon\omega}\cos(\omega k L) \ . 
\label{sigmapss1}
\end{eqnarray}
Note that keeping the convergence factor with the $\epsilon$ is crucial to induce the correct contour in the integration. We can integrate the spectral function \eqref{sigmapss1} to check that we recover the Casimir energy
\begin{eqnarray}
\rho_{ren}&=&\int_{0}^{+\infty}\dfrac{d\omega}{2\pi}2\omega\sum_{k=1}^{+\infty}e^{-\epsilon\omega}\cos(\omega L k)\nonumber \\ &=&\dfrac{1}{\pi L^2}\sum_{k=1}^{+\infty}\dfrac{\epsilon^2-k^2}{(\epsilon^2+k^2)^2} 
\rightarrow -\dfrac{\pi}{6L^2} \ . \label{rhos1}
\end{eqnarray}

Using the Fourier series representation of the periodic repetition of the delta function one can show that the sum \eqref{sigmapss1}, with $\epsilon\rightarrow 0$,  is equivalent to \eqref{sigmas1}. Thus, using point-splitting we recover the spectral function obtained in the previous subsection using a weight function.

\subsection{Spectral function or spectral distribution?}
The natural question is: does $\sigma(\omega)$  carry some extra physical content, as compared to the energy density? Or in other words: is the spectral function $\sigma(\omega)$ meaningful only when integrated over the whole frequency range, or can one assign a physical value to an integration over an interval, $\int_{\omega_0}^{\omega_0+\Delta\omega}\sigma(\omega)d\omega$ ? 

To answer this question we can integrate $\sigma(\omega)$ in~\eqref{sigmapss1} convoluted with some other functions. For example we can take functions which are non-zero only in regions with only one delta function contribution. For example if we choose, for $j\in \mathbb{N}$
\begin{equation}
f_j(\omega)=\left\{\begin{array}{ll} 1 &\hspace{5mm}, (2\pi j-\pi)/L <\omega<(2\pi j+\pi)/L \\ 0 &\hspace{5mm},  \mathrm{otherwise}\end{array}\right. \ , 
\end{equation}
then, using \eqref{sigmapss1},
\begin{eqnarray}
\int_{0}^{+\infty}f_j(\omega)\sigma(\omega)\dfrac{d\omega}{2\pi}&=& 0 \ .
\end{eqnarray}
This is exactly what we obtain from integrating $f_j(\omega)$ with the delta distribution centred at $2\pi j/L$, and subtracting the result from integrating $f_j(\omega)$ with $\omega/(2\pi)$, cf.~\eqref{sigmas1}. But integration over the first (remaining) interval yields
\begin{eqnarray}
\int_{0}^{\pi/L}\sigma(\omega)\dfrac{d\omega}{2\pi}=-\frac{\pi}{4L^2} \ , 
\end{eqnarray}
rather than $-\pi/6L^2$, which is the renormalised energy density. This is a consequence of manipulating divergent infinite sums without regularising them. Of course, if one keeps the convergence factor $e^{-\epsilon\omega}$ in \eqref{rhos1}, and integrates again over the full range of frequencies, one will obtain the correct result. This makes clear that the spectral function $\sigma(\omega)$ is only meaningful when regularised.

What if we include a regulator in $\sigma(\omega)$? Does the spectral function then acquire a physical meaning when integrated over  a compact range of frequencies (as opposed to all frequencies)? A hint to answer this question comes from work on semi-transparent boundaries \cite{Jaekel:1991j}. Therein, the Casimir force is given by a difference of pressures inside the plates and outside the plates, computed from scattering of modes inside and outside the plates. In such computation there is a physical coefficient playing the role of a reflection factor, which becomes the physical regulator. A related idea was discussed in \cite{PhysRevA.48.2962}.

\begin{figure*}
\begin{center}
\begin{picture}(0,80)
\put(0,80){\scriptsize $L^2 F_1(100,\omega_0)$}
\put(250,80){\scriptsize $L^2 F_1(1000,\omega_0)$}
\end{picture}
\begin{tabular}{cc}
 \hspace{-4mm} \includegraphics[width=0.45\textwidth]{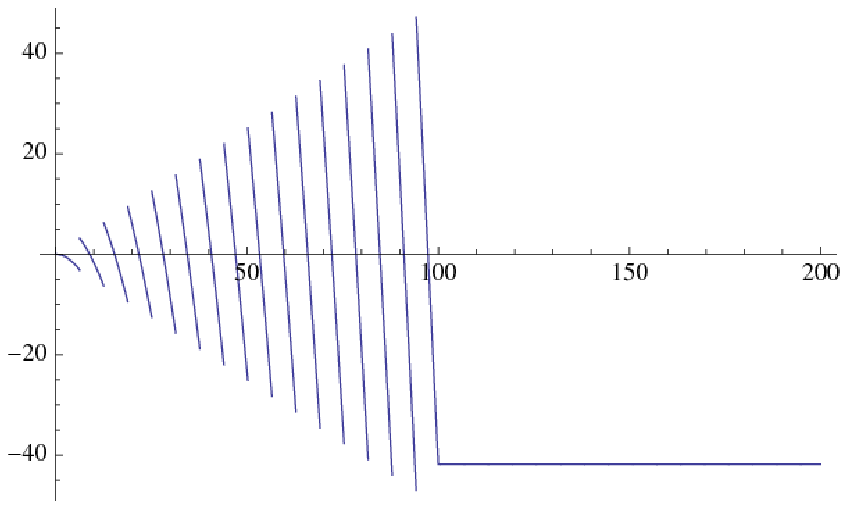} \ \ \ \ \  &
  \includegraphics[width=0.45\textwidth]{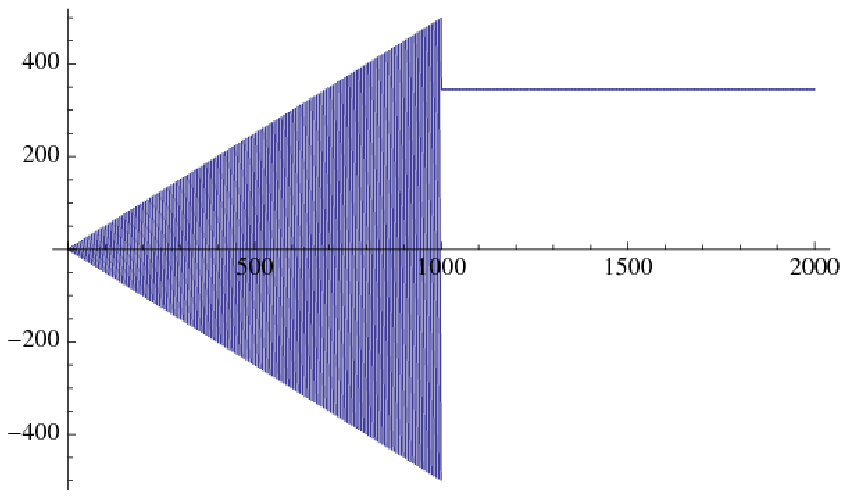}
\end{tabular}
\end{center}
   \caption{\label{Plots_intSpect_theta} Integral of the spectral function regularised with a step function at $q$, up to an energy $\omega_0$  {\em Left:} $F_1(100,\omega_0)$,  {\em Right:} $F_1(1000,\omega_0)$.} 
\end{figure*}

Assuming that the spectral function has a physical meaning, we could think about integrating it over a finite range of frequencies, to mimic the situation of transparency in the remaining range of frequencies. Let us first consider this is done in a sharp way, i.e. by introducing a Heaviside step function $H(x)$ cut-off: 
\begin{equation}
F_1(q,\omega_0)\equiv \int_0^{\omega_0} \dfrac{d\omega}{2\pi}\sigma(\omega)H(q-\omega) \ .
\end{equation} 
In Fig.~\ref{Plots_intSpect_theta} we plot $F_1(q,\omega_0)$. It oscillates wildly and it does not converge to the renormalised energy density in the limit $\omega_0\rightarrow \infty$. 

However, we can introduce a smoother regulator, as we would expect in an actual physical situation such as in~\cite{Jaekel:1991j}. Using the exponential cut-off $e^{-\epsilon \omega}$, so that we can interpret $q=1/\epsilon$ as the scale at which the theory is UV completed; then consider
\begin{equation}
F_2(q,\omega_0)\equiv \int_0^{\omega_0} \dfrac{d\omega}{2\pi}\sigma(\omega)e^{-\omega/q} \ .
\end{equation}

\begin{figure*}
\begin{center}
\begin{picture}(0,80)
\put(0,75){\scriptsize $L^2 F_2(100,\omega_0)$}
\put(173,1){\vector(1,0){70}}
\end{picture}
\begin{tabular}{cc}
 \hspace{-4mm} \includegraphics[width=0.45\textwidth]{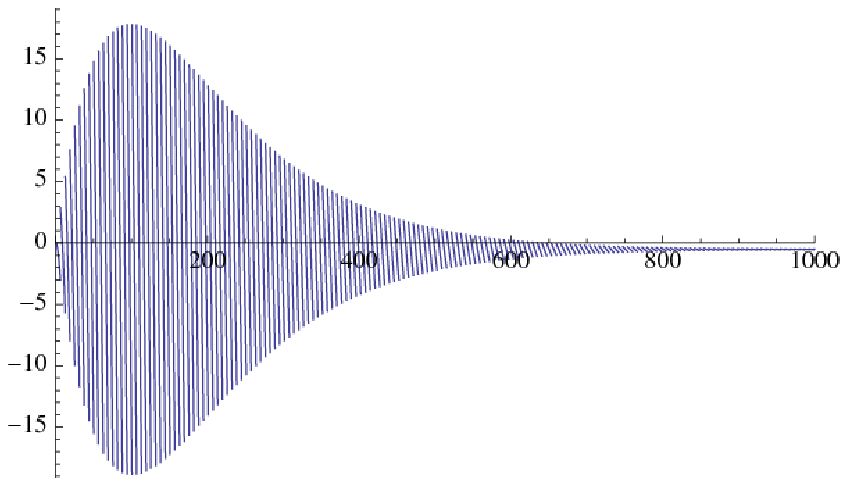} \ \ \ \ \  &
  \includegraphics[width=0.45\textwidth]{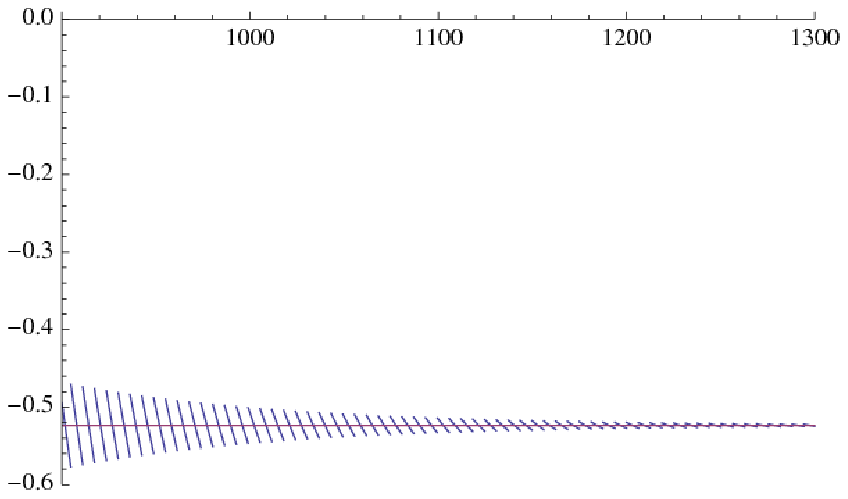}
\end{tabular}

\vspace{1cm}

\begin{picture}(0,80)
\put(0,75){\scriptsize $L^2 F_2(1000,\omega_0)$}
\put(173,1){\vector(1,0){70}}
\end{picture}
\begin{tabular}{cc}
 \hspace{-4mm} \includegraphics[width=0.45\textwidth]{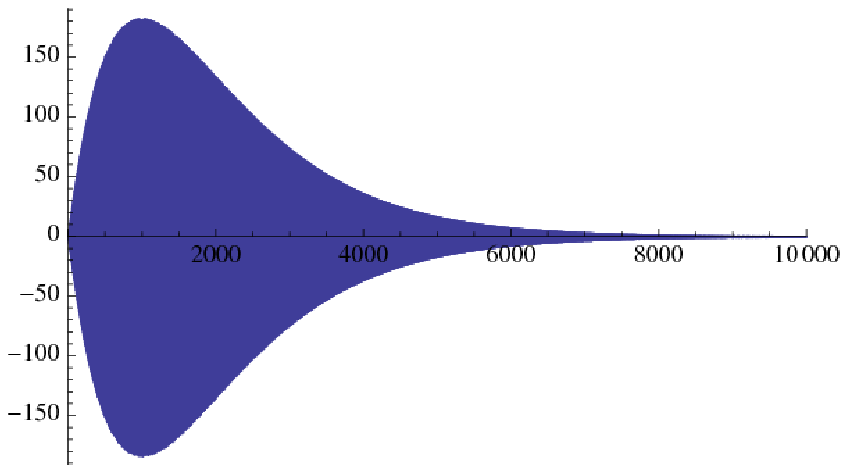} \ \ \ \ \  &
  \includegraphics[width=0.45\textwidth]{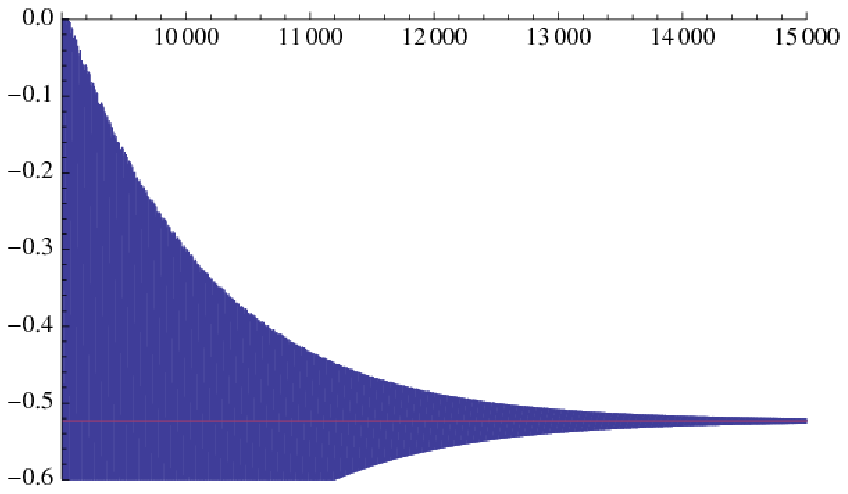}
\end{tabular}
\end{center}

   \caption{\label{Plots_intSpect} Integral up to an energy $\omega_0$, of the spectral function regularised with an exponentially decaying function  {\em Left:} $F_2(100,\omega_0)$,  {\em Right:} $F_2(1000,\omega_0)$. The asymptotic result is indicated by the horizontal line in the zoomed plots on the right.}
\end{figure*}
This function is plotted in Fig.~\ref{Plots_intSpect}. It can be seen that the integrated value of the spectral function now approaches asymptotically the Casimir energy density of the system. Thus, for the spectral function to be physical, a regulator that suppresses the UV frequencies is crucial and it should correspond to an actual physical cut-off of the theory. 
It is a remarkable fact that the Casimir force does not depend on the details of such regulator, but only on the existence of a high energy ``smooth enough'' damping of the modes.

\subsection{Interpretation of $\sigma(\omega)$ and computational shortcut}\label{shortcut}
Inspired by the previous example, and since the definition of the spectral density \eqref{sf} is not derived from first principles, let us reconsider it starting from the (bare) energy density formula.  Keeping the convergence factor~$e^{-\epsilon \omega_n}$
\begin{eqnarray}
\rho &=& \dfrac{1}{2L}\sum_{n=-\infty}^{+\infty}\omega_n e^{-\epsilon \omega_n} \nonumber\\
     &=&\int_{-\infty}^{+\infty} d\omega \omega\dfrac{1}{2L}\sum_{n=-\infty}^{+\infty}\delta(\omega-\omega_n)e^{-\epsilon \omega_n} \nonumber \\
     &=&\int_{0}^{+\infty} d\omega \frac{\omega}{L}\sum_{n=1}^{+\infty}\left(\delta(\omega-\omega_n)+\delta(\omega+\omega_n)\right)e^{-\epsilon \omega_n} . \hspace{8mm} 
\end{eqnarray}
Observe that the second delta function does not contribute. From this we see that the bare spectral function with the normalisation \eqref{maindef} is
\begin{equation}
\sigma_{\mathrm{bare}} = \dfrac{2\pi}{L}\sum_{n=1}^{+\infty}\omega_n\left(\delta(\omega-\omega_n)+\delta(\omega+\omega_n)\right)e^{-\epsilon \omega_n} \ .
\end{equation}
If we replace here the Fourier expansion of the delta function
\begin{eqnarray}
\sigma_{\mathrm{bare}} &=& \sum_{n=1}^{+\infty}\int_{-\infty}^{\infty}\dfrac{d\tau}{L}\omega_n \left[e^{i\tau(\omega-\omega_n)}+e^{i\tau(\omega+\omega_n)}\right]e^{-\epsilon \omega_n}\nonumber \\
&=& \dfrac{2}{L}\int_{-\infty}^{\infty}d\tau e^{i\tau\omega}\sum_{n=1}^{+\infty}\omega_n\cos(\tau\omega_n)e^{-\epsilon \omega_n}\nonumber \\
&\stackrel{\eqref{ps}}{=}& 2\int_{-\infty}^{\infty} \langle 0| T_{00}(\tau)|0 \rangle e^{i\tau\omega} d\tau \; .
\end{eqnarray}
So the fact that the spectral function is the Fourier transform of the point split energy density is \textit{not restricted to using point splitting regularisation}; indeed we have not used it in this argument. This expression appears simply  from Fourier expanding the delta functions.

The lesson from this example is that the renormalised spectral density $\sigma(\omega)$ is simply the difference between the bare spectral density above and the spectral density of the reference background. In Appendix A we show that this simple `rule of thumb' yields the correct result for the computation of the spectral density in another well known example: a massless scalar field in $\mathbb{R}^3\times S^1$.

\section{Application to $\mathbb{R}\times S^3$}\label{sec:fields}
We will now ask another question concerning the physical content of the spectral function: does it carry any hint concerning the sign of the Casimir force? It is well known that the Casimir force can either be attractive or repulsive depending on details of the geometry, topology or boundary conditions that originates it (see \cite{Bordag:2001qi} for a review). What determines the sign of the Casimir force is still an open problem.
 
This question is particularly relevant since Casimir forces are a dominant interaction at the nanometer scale and therefore of relevance for the construction of nano-devices. Repulsive Casimir forces (see e.g. \cite{Kenneth:2002ij,Rosa:2010wp}) could allow quantum levitation and therefore avoid collapse and permanent adhesion between nearby surfaces, which is a problem for microelectromechanical systems. Moreover, repulsive Casimir forces are not just a theoretical possibility since there is some experimental evidence \cite{Milling,Lee,Munday}.

In order to tackle this question we shall consider an example where it has been shown that the effect of the Casimir energy can change from attractive to repulsive, when one varies the parameters of the system: a scalar field with a mass parameter and a coupling to the Ricci scalar in the Einstein Static Universe (ESU)~\cite{Herdeiro:2005zj,Herdeiro:2007eb}.

\subsection{Setup}
We consider a scalar field theory in the $3+1$ dimensional ESU
\begin{equation}
ds^2_{ESU}=-dt^2+R^2d\Omega_{S^3} \ . \label{metric}
\end{equation} 
The Ricci scalar of the geometry is $\mathcal{R}=6/R^2$, where the radius of the universe is $R$. This geometry is perturbed by the quantum  vacuum fluctuations of a real scalar field with action
\begin{equation}\label{scalar_action}
\mathcal{S}_{\Phi}=\int d^{4}x \sqrt{-g}\left(-\frac{1}{2}\partial_{\alpha}\Phi\partial^{\alpha}\Phi-\frac{1}{2}\mu^2\Phi^2-\frac{1}{2}\xi \mathcal{R}\Phi^2\right) \; .
\end{equation}
We are interested in analysing the energy spectrum of the quantum field in a vacuum state (with respect to the global timelike Killing vector). This can be done by solving the classical equations of motion for the scalar field, then quantising the theory and finally computing expectation values of observables in such a state. Since we are interested in the energy density we have to look at the expectation value of the energy momentum tensor
\begin{multline}
T_{\mu
  \nu}^{\Phi}=\partial_{\mu}\Phi\partial_{\nu}\Phi+\xi\left(\mathcal{R}_{\mu\nu}-D_{\mu}D_{\nu}\right)\Phi^2+\\+g_{\mu\nu}\left(2\xi-\frac{1}{2}\right)\left[\partial_{\alpha}\Phi\partial^{\alpha}\Phi+(\mu^2+\xi\mathcal{R})\Phi^2\right] \ . \label{EM_tensor} \end{multline}  

This procedure was explained in detail in~\cite{Herdeiro:2005zj,Herdeiro:2007eb}, so here we simply summarise the main results needed for our study in $3+1$ dimensions. The scalar field is expanded in the Heisenberg picture using the eigenmodes
\begin{equation}
\phi_\alpha(t,\underline{x})=\dfrac{1}{\sqrt{2\omega_\ell}}e^{-i\omega_\ell
t}Y_\alpha(\underline{x})\label{killing_eigen} \ , 
\end{equation}
where $\alpha= \{\ell,m_R,m_L\} : \ell \in \mathbb{N}_0 \wedge -\ell <m_L,m_R< \ell \in \mathbb{Z}$, $Y_\alpha(\underline{x})$ are the hyperspherical harmonics and $\underline{x}$ denotes the angular coordinates on $S^3$. The eigenfrequencies are
\begin{equation}
\omega_{\ell}=\sqrt{k^2+a^2} \; , 
\label{frespe2}
\end{equation}
with
\begin{equation}
k\equiv \frac{\ell+1}{R} \ ,\ \   a^2\equiv \mu^2+
\frac{6\xi-1}{R^2} 
\ . \label{defi}\end{equation}
The Vacuum Expectation Value (VEV) of any component of $T_{\mu\nu}^{\Phi}$ can be expressed in terms of ${\left<0\right|(T^{\Phi})^{0}}_{0}\left|0\right>$ using the conservation of energy. Its bare value is formally divergent, so it requires a regularisation procedure followed by renormalisation which we address in the next section.

\subsection{The spectrum on the ESU from a generic regularisation}

\begin{figure*}
\begin{center}
\begin{picture}(0,60)
\put(0,60){\scriptsize $F(1000,1,\omega_0)$}
\put(173,1){\vector(1,0){35}}
\end{picture}
\begin{tabular}{cc}
 \hspace{-4mm} \includegraphics[width=0.38\textwidth]{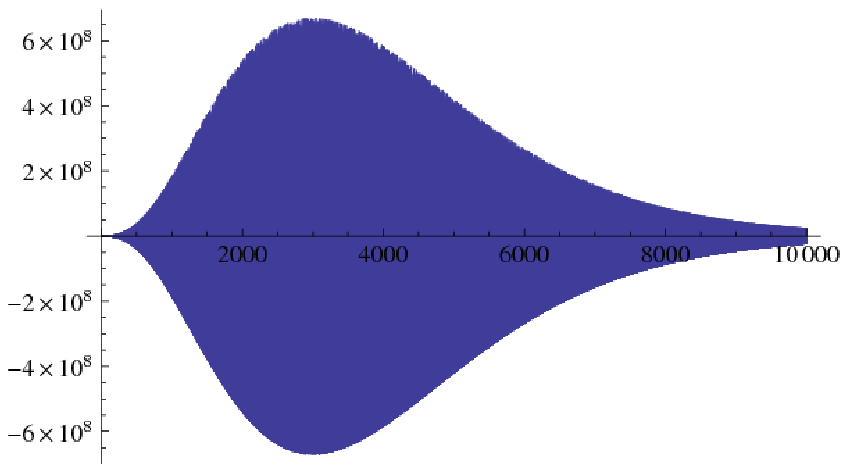} \ \ \ \ \  &
  \includegraphics[width=0.4\textwidth]{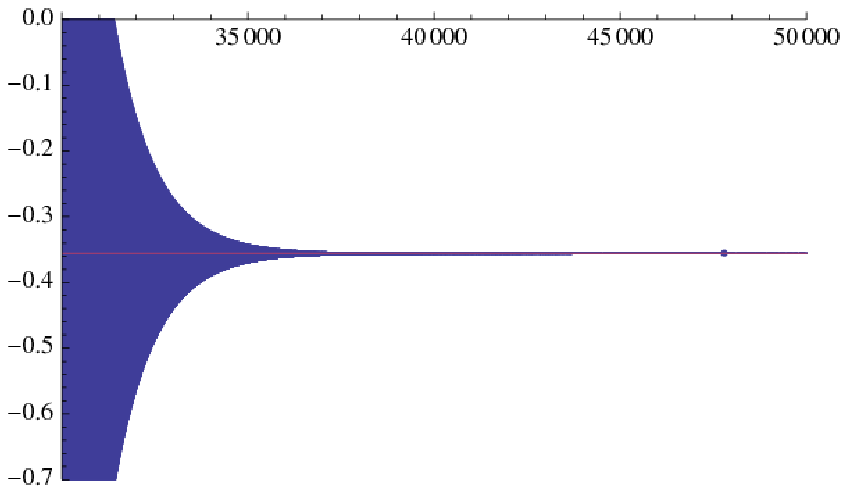}
\end{tabular}

\vspace{0.5cm}
\begin{picture}(0,60)
\put(0,60){\scriptsize $F(1000,-0.9^2,\omega_0)$}
\put(173,1){\vector(1,0){35}}
\end{picture}
\begin{tabular}{cc}
 \hspace{-4mm} \includegraphics[width=0.38\textwidth]{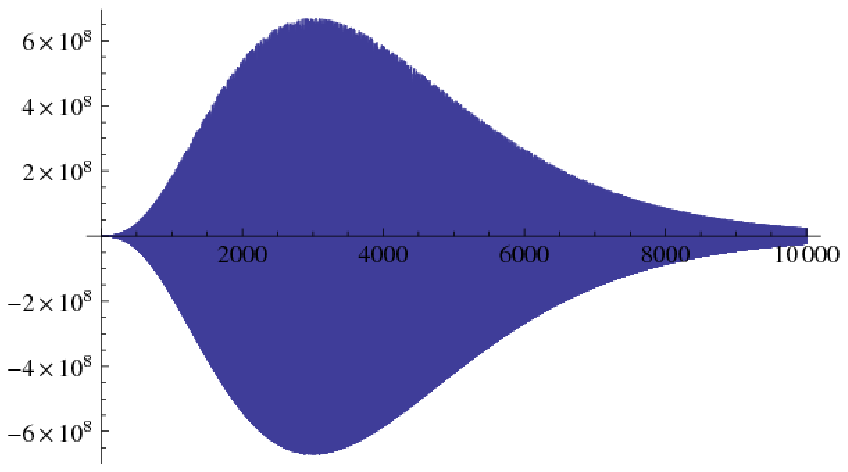} \ \ \ \ \  &
  \includegraphics[width=0.4\textwidth]{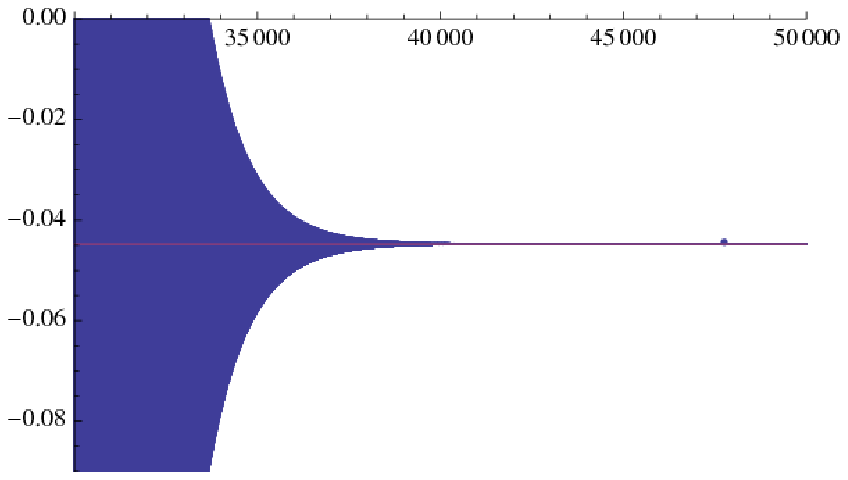}
\end{tabular}

\vspace{0.5cm}
\begin{picture}(0,60)
\put(0,60){\scriptsize $F(1000,-a_0^2,\omega_0)$}
\put(173,1){\vector(1,0){35}}
\end{picture}
\begin{tabular}{cc}
 \hspace{-4mm} \includegraphics[width=0.38\textwidth]{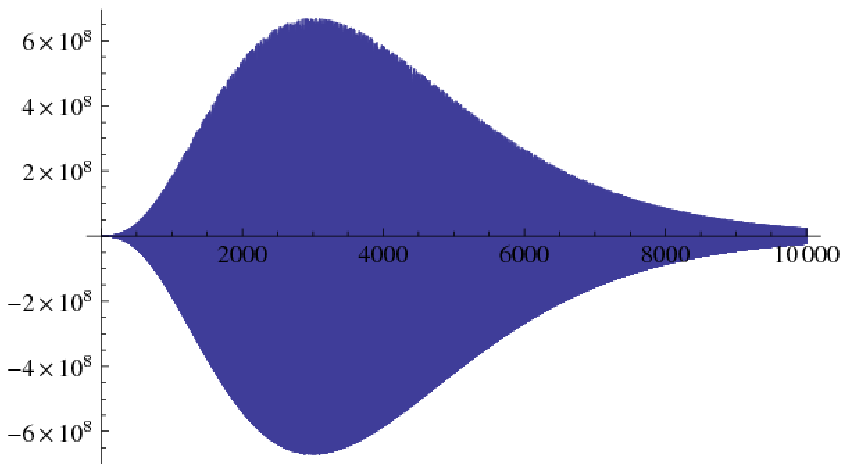} \ \ \ \ \  &
  \includegraphics[width=0.4\textwidth]{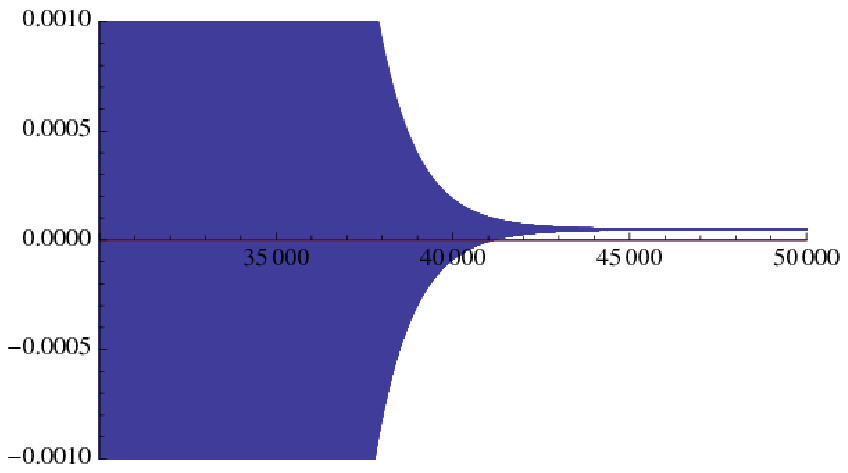}
\end{tabular}

\vspace{0.5cm}  
\begin{picture}(0,60)
\put(0,60){\scriptsize $F(1000,-0.7^2,\omega_0)$}
\put(173,1){\vector(1,0){35}}
\end{picture}
\begin{tabular}{cc}
 \hspace{-4mm} \includegraphics[width=0.38\textwidth]{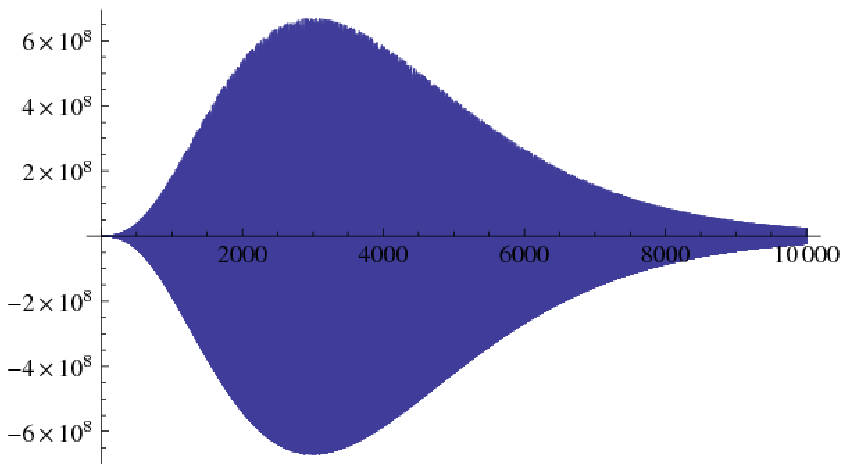} \ \ \ \ \  &
  \includegraphics[width=0.4\textwidth]{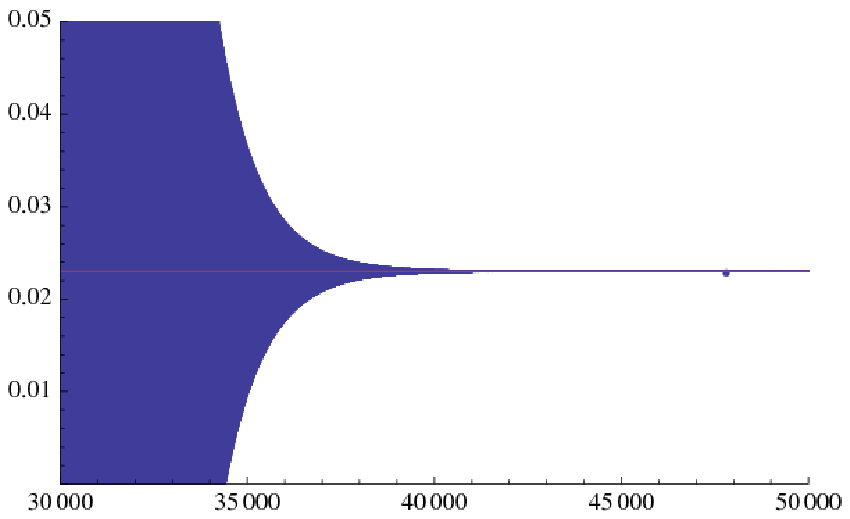}
\end{tabular}

\vspace{0.5cm}  
\begin{picture}(0,60)
\put(0,60){\scriptsize $F(1000,0,\omega_0)$}
\put(173,1){\vector(1,0){35}}
\end{picture}
\begin{tabular}{cc}
 \hspace{-4mm} \includegraphics[width=0.38\textwidth]{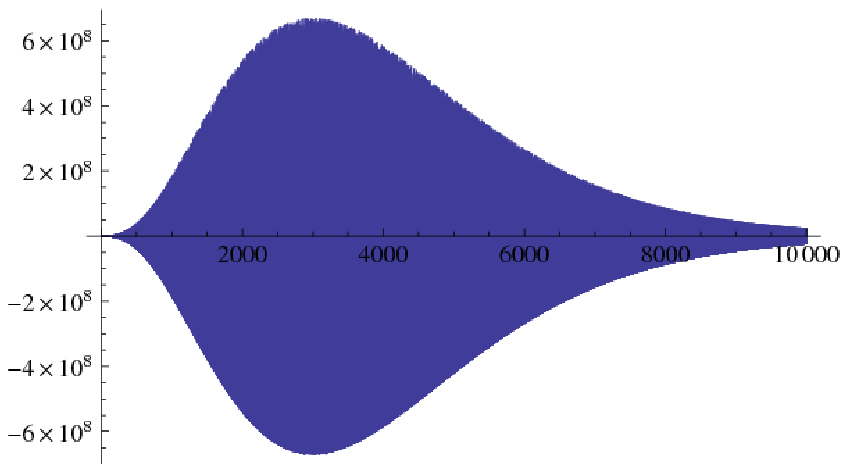} \ \ \ \ \  &
  \includegraphics[width=0.4\textwidth]{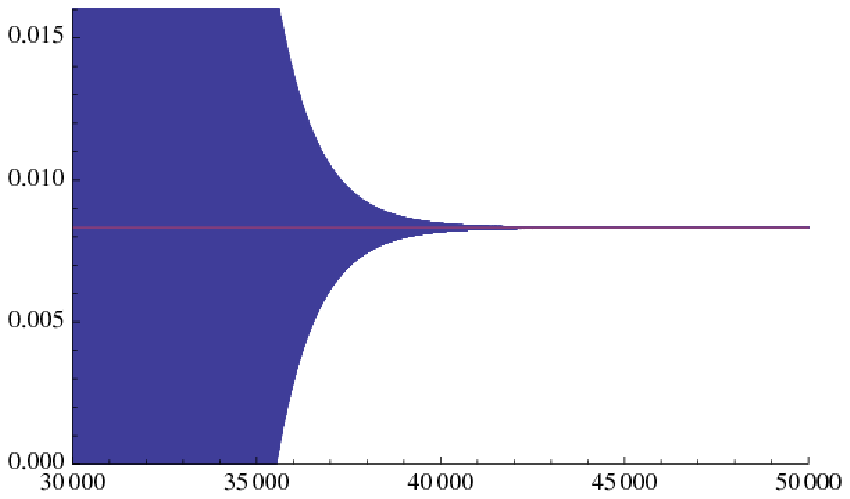}
\end{tabular}
\end{center}
   \caption{\label{Plots_intSpectRS3} Integral up to energy $\omega_0$, of the spectral function regularised with an exponentially decaying function, for various choices of $a^2$. The asymptotic result is indicated by the horizontal line in the zoomed plots (right). Note that $-a_0^2$ is the value for which the renormalised energy density vanishes; which can be seen on the corresponding zoomed plot up to an error of~$10^{-4}$.}
\end{figure*}

As seen in~\cite{Herdeiro:2007eb}, the renormalisation of the energy momentum tensor on the ESU can be performed just by specifying some generic physical properties for the UV completion of the theory, encoded in a generic regulator. Then the regularised energy density is simply
\begin{equation}
\rho_{0}=\frac{1}{2V}\sum_{\ell=0}^{+\infty}d_{\ell}\omega_{\ell} \, g\left(\gamma L \omega_\ell\right)  \label{reg_general}\ ,
\end{equation} 
where $V$ is the volume of $S^3$, $g$ is a generic regularisation function that goes to $1$ in the infrared, $\gamma \ll 1 $ is the regularisation parameter, $L$ is the typical length scale of the system, and the degeneracy of the modes is
\begin{equation}
d_{\ell}=(\ell+1)^2 \ .
\end{equation}
Using the Abel-Plana formula $\rho_0$ can be re-written as 
\begin{equation}
 \rho_0=\dfrac{R^3}{2V}\int_{\omega_{min}}^{+\infty}d\omega \sqrt{\omega^2-a^2}\omega^2g(\gamma L \omega)+\rho_{ren} \ ,
\end{equation} 
where $\omega_{min}=a$ if $a^2>0$, otherwise $\omega_{min}=0$. The first integral is the quantity that is discarded and contains all the divergences that renormalise various gravitational couplings. In fact, to obtain the spectrum, this is exactly the quantity that we need to subtract from the discrete spectrum. This  can be written as
\begin{equation}
 \rho_{div}=\int_{0}^{+\infty}\dfrac{d\omega}{2\pi} \dfrac{\pi R^3}{V}\sqrt{\omega^2-a^2}\omega^2g(\gamma L \omega)\theta(\omega-\omega_{min}) \ .
\end{equation}
So now, using our `rule of thumb' we can just subtract the integrand from the discrete spectrum to obtain the spectrum
\begin{multline}
\sigma_{ESU}=\frac{\pi\omega}{V}\left[R^2(\omega^2-a^2)\sum_{\ell=0}^{+\infty} \delta\left(\omega-\omega_\ell\right)-\right.\\\left.-R^3\omega\sqrt{\omega^2-a^2}\theta(\omega-\omega_{min}) \right]g(\gamma L\omega) \ .
\end{multline}
Once again we can define the integral of the spectrum up to an energy $\omega_0$ and make the plot for various $a$. If for example we take the massless case, then we can show that for this case, the quantity $\rho+3p$, which determines if the strong energy condition is obeyed, simplifies to
\begin{equation}
\rho+3p=2\rho \ ,
\end{equation}
so that it is simply the sign of $\rho$ that determines if the gravitational effect of the Casimir energy is attractive ($\rho>0$) or repulsive ($\rho<0$). The change from a repulsive to an attractive effect can be obtained by varying $a^2R^2$, or equivalently $\xi$ ~\cite{Herdeiro:2005zj,Herdeiro:2007eb}. For convenience we define the integrated spectrum
\begin{equation}
F(q,a^2,\omega_0)=2R^4V^{(3)}\int_0^{\omega_0}\sigma_{ESU}(\omega)\dfrac{d\omega}{2\pi} \ ,
\end{equation}
with $g(\gamma L\omega)=e^{-\omega/q}$. Some examples are exhibited in Fig.~\ref{Plots_intSpectRS3}, for which one varies $a^2$ such that the effect varies from attractive to repulsive. The conclusion is that there is \textit{no manifest qualitative change in the spectrum} and the variation in character of the Casimir force can only be observed upon integration of the spectral function. Thus, it seems that again, the Casimir spectrum carries no added value, relatively to the Casimir energy density, in terms of physical content.

\section{Concluding remarks}\label{sec:Conclusions}
In this paper we have reconsidered the Casimir spectrum, defined in~\cite{Ford:1988gt}. Based on the study of three different examples (two of which had already been considered in~\cite{Ford:1988gt}) we have reached the following conclusions: 1) despite the apparent regularisation scheme dependent definition \eqref{sf}, the Casimir spectrum is mathematically well defined, in the sense of being scheme independent; 2) a simple interpretation for the spectrum is provided: the renormalised spectral density $\sigma(\omega)$ is simply the difference between the bare spectral density and the spectral density of the reference background. This provides us with a simple rule to write down a formal expression for the spectrum; 3) the spectrum should, however, be faced as a distribution, rather than a function. That is, it is not differentiable but it is integrable, and also it is only physically meaningful when integrated over a sufficiently large range of frequencies, with a high energy smooth enough regulator. Thus we find no evidence that the spectrum carries an added value, in terms of physical content, relatively to the energy density itself. In particular, by considering an example where the effect of the Casimir energy can vary its character from repulsive to attractive, we found that there is no manifest imprint of this change of character in the properties of the spectrum. We note that a possible relation between the Casimir spectrum and the sign of the Casimir force was previously discussed in~\cite{PhysRevA.48.2962}.


\section*{Acknowledgements}
We would like to thank L. Ford for comments on a draft of this paper and Gon\c calo Dias for collaboration during the early stages of this work. We would also like to thank Manuel Donaire and Filipe Paccetti for discussions. M.S. is supported by the FCT grant SFRH/BPD/69971/2010. J.E.S. would like to thank the MPIPkS for support in the framework of its Visitors Program. This work is also supported by the grant CERN/FP/109306/2009.


\section*{Appendices}
\appendix

\section{Check for $S^1\times \mathbb{R}^3$} \label{Check_Ford}

In~\cite{Ford:1988gt} the spectrum of the Casimir effect, for a massless scalar field in flat spacetime, with one periodic spatial direction was computed using Green's functions. In this appendix we show that the shortcut suggested in Section \ref{shortcut} yields the same result, which is also checked by using point-splitting regularisation. 

In this case, due to the boundary conditions, the field expansion is
\begin{equation}
\Phi = \int d^2k\sum_\ell \left[a_\ell(k)\phi_\ell(k,x^\mu)+a_\ell^\dagger(k)\phi_\ell^*(k,x^\mu)\right] \ ,
\end{equation}
where the creation and annihilation operators obey standard harmonic oscillator commutation relations, and
\begin{equation}
\phi_\ell(k,x^\mu)=\dfrac{1}{2\pi\sqrt{2\omega_{\ell,k} L}}e^{-i\omega_{\ell,k} t+i\left(\frac{2\pi \ell}{L}x+k_y y + k_z z\right)} \ , 
\end{equation} 
with
\begin{equation}
\omega_{\ell,k}=\sqrt{\left(\frac{2\pi \ell}{L}\right)^2+k^2} \ .
\end{equation}

The first step is to compute the vacuum expectation value of the regularised energy density from the energy momentum tensor
\begin{widetext}
\begin{multline}
\left<0\right|T_{00}(x-x')\left|0\right>\equiv \dfrac{1}{2}\left<0\right|\partial_0\Phi\partial_{0'}\Phi'+\dfrac{1}{2}\partial_{\alpha'}\Phi'\partial^{\alpha}\Phi+x\leftrightarrow x'\left|0\right>\\=  \dfrac{1}{2 L}\int \dfrac{d^2k}{(2\pi)^2}\sum_{\ell=-\infty}^{+\infty} \omega_{\ell,k}\cos\left[\omega_{\ell,k}(t-t')-\frac{2\pi\ell}{L}(x-x')\right]e^{i(k_y(y-y')+k_z(z-z'))} \ , 
\end{multline}
\end{widetext}
where we have used the field expansion and the canonical commutation relations for the creation and annihilation operators. 

Taking time separation only $\tau=t-t'$, and introducing a convergence factor to make sure the sum converges we have
\begin{equation}
\left<0\right|T_{00}(\tau)\left|0\right>=  \dfrac{1}{2 L}\int \dfrac{d^2k}{(2\pi)^2}\sum_{\ell=-\infty}^{+\infty} \omega_{\ell,k}\cos\left(\omega_{\ell,k}\tau\right)e^{-\epsilon \omega_n} \ . \label{a5}
\end{equation}

Before applying the Abel-Plana formula let us use the insight of Section \ref{shortcut} to anticipate the answer (since the spectrum was seen as a difference between the bare spectral density and the flat space spectral density). The bare spectral density is obtained by inserting a delta function
\begin{eqnarray}
\rho&=&  \dfrac{1}{2 L}\int \dfrac{d^2k}{(2\pi)^2}\sum_{\ell=-\infty}^{+\infty} \omega_{\ell,k} \\
&=&  \int \dfrac{d\omega}{2\pi}\dfrac{1}{2 L}\int \dfrac{d^2k}{(2\pi)^2}\sum_{\ell=-\infty}^{+\infty} \omega_{\ell,k}2\pi\delta(\omega-\omega_{\ell,k}) \; . \nonumber
\end{eqnarray}
So
\begin{eqnarray}
\sigma_{bare}(\omega)&=&\dfrac{1}{2 L}\int \dfrac{d^2k}{(2\pi)^2}\sum_{\ell=-\infty}^{+\infty} \omega_{\ell,k}2\pi\delta(\omega-\omega_{\ell,k})\nonumber\\
&=&\dfrac{1}{2 L}\sum_{\ell=-\infty}^{+\infty}\int_0^{+\infty} dk k \omega_{\ell,k}\delta(\omega-\omega_{\ell,k})
\nonumber \\
  &=&
\dfrac{\omega^2}{2 L}\left(1+2\sum_{\ell=1}^{+\infty}\theta(\omega L-2\pi |\ell|)\right) \ .\label{eqAsigmabare}
\end{eqnarray}
The flat limit spectrum is clearly
\begin{equation}
\sigma_{flat}(\omega)=\dfrac{\omega^3}{2\pi } \ , \label{eqAsigmainf}
\end{equation}
so we expect the spectrum to be the difference between~\eqref{eqAsigmabare} and~\eqref{eqAsigmainf}:
\begin{equation}
\sigma(\omega)
=\dfrac{\omega^2}{2\pi L}\left(\pi-\omega L+2\pi\sum_{\ell=1}^{+\infty}\theta(\omega L-2\pi |\ell|)\right) \; . \label{expected}
\end{equation}
The function in parenthesis is just a (periodic) saw function, which is a repetition of the first linear segment $\pi-\omega L$. This is exactly the same as the result in~\cite{Ford:1988gt} (except that the result therein has an incorrect overall minus sign). Note that we have not included a converging factor $e^{-\epsilon \omega}$, which should be implicit for this distribution to make sense.

\begin{figure}[b]
\begin{picture}(0,150)
\put(15,140){\scriptsize $L \sigma(\omega)$}
\end{picture}
  \centering\includegraphics[width=0.45\textwidth]{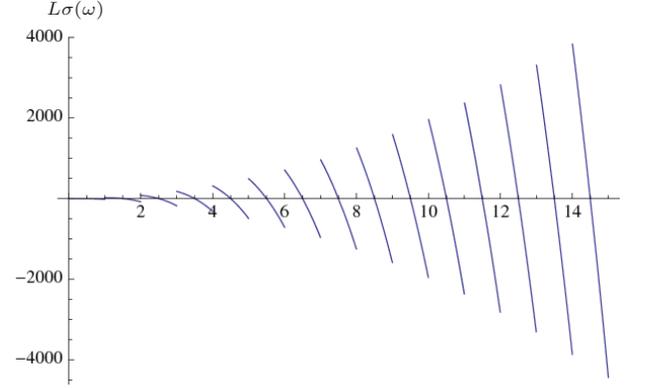}
   \caption{\label{Plot_PS_S1R3}  Spectral function from point splitting for $S^1\times\mathbb{R}^3$ .} 
   \label{fig2}
\end{figure}

Once again this result can also be obtained by applying the Abel-Plana formula to~\eqref{a5} and performing manipulations similar to those in the appendix~B of~\cite{Herdeiro:2007eb}. Start with the point split energy density
\begin{equation}
\rho_\tau=\dfrac{1}{2L}\int \dfrac{d^2k}{(2\pi)^2}\sum_{\ell=-\infty}^{+\infty}\omega_{\ell,k}\cos(\omega_{\ell,k}\tau)e^{-\epsilon \omega_{\ell,k}} \ ,
\end{equation}
and apply the Abel-Plana formula to get
\begin{widetext}
\begin{equation}
\rho_\tau\equiv\left<0\right|T_{00}(\tau)\left|0\right>=\int \dfrac{d^3k}{(2\pi)^3} k\cos\left(k\tau\right)e^{-\epsilon k}+\dfrac{4\pi}{L^4}\int d^2q\int_q^{+\infty}dx \dfrac{\sqrt{x^2-q^2}}{e^{2\pi x}-1}\cosh\left(2\pi\sqrt{x^2-q^2}\frac{\tau}{L}\right)\cos\left(2\pi\frac{\epsilon}{L}\sqrt{x^2-q^2}\right) \ .
\end{equation}
\end{widetext}
 The first (divergent) term is a background independent constant, so it renormalises the cosmological constant.  Since we are in flat space and all other curvature invariants are zero it can be shown that no further couplings get renormalised. 
The renormalised point split energy density is then (we have performed the angular integration in $d^2q$)
\begin{widetext}
\begin{multline}
\rho_{\tau,ren}=\dfrac{2\pi^2}{L^4}\int_0^{+\infty} dq\, q\int_q^{+\infty}dx \dfrac{\sqrt{x^2-q^2}}{e^{2\pi x}-1}\left[e^{-2\pi\sqrt{x^2-q^2}\frac{\tau+i\epsilon}{L}}+e^{-2\pi\sqrt{x^2-q^2}\frac{\tau-i\epsilon}{L}}+\right. \\ \left.+e^{2\pi\sqrt{x^2-q^2}\frac{\tau-i\epsilon}{L}}+e^{2\pi\sqrt{x^2-q^2}\frac{\tau+i\epsilon}{L}}\right] \ .
\end{multline}
\end{widetext}
If we expand the denominator in a geometric series once again, and exchange the order of integration, then we can perform the integrals explicitly to give
\begin{multline}
\rho_{\tau,ren}=\dfrac{1}{(2\pi)^2L^4}\sum_{j=1}^{+\infty}\dfrac{1}{j}\left[\dfrac{1}{(\frac{\tau-i\epsilon}{L}-j)^3}-\dfrac{1}{(\frac{\tau-i\epsilon}{L}+j)^3}+\right.\\\left.+\dfrac{1}{(\frac{\tau+i\epsilon}{L}-j)^3}-\dfrac{1}{(\frac{\tau+i\epsilon}{L}+j)^3}\right] \ .
\end{multline}
Once again if we take the Fourier transform, we can perform the integral by closing in the upper complex semi-plane and reduce the problem to the calculation of residues. The result is then
\begin{equation}
\sigma(\omega)=\dfrac{\omega^2}{2\pi L}\sum_{j=1}^{+\infty}\dfrac{2\sin(\omega L j)}{j}e^{-\epsilon \omega} \ , 
\end{equation}
which we can check, by considering the Fourier series of a saw-tooth function, to be exactly the same as the distribution~\eqref{expected} anticipated above - Fig. \ref{fig2}.

\bibliographystyle{h-physrev4}
\bibliography{SpectrumESU}

\end{document}